\documentclass[10pt,twoside,twocolumn]{article}
\usepackage{times,mathptmx}
\usepackage{amssymb}
\usepackage{balance} 

\usepackage{graphicx} 
\usepackage{lastpage}
\usepackage[format=plain,justification=raggedright,singlelinecheck=false,font=small,labelfont=bf,labelsep=space]{caption} 
\usepackage{fancyhdr}

\begin{document}

\thispagestyle{plain}
\renewcommand{\thefootnote}{\fnsymbol{footnote}}
\renewcommand\footnoterule{\vspace*{1pt}%
\hrule width 3.4in height 0.4pt \vspace*{5pt}} 
\setcounter{secnumdepth}{5}

\renewcommand{\figurename}{\small{Fig.}~}

\renewcommand{\headrulewidth}{1pt} 
\renewcommand{\footrulewidth}{1pt}
\setlength{\arrayrulewidth}{1pt}
\setlength{\columnsep}{6.5mm}

\twocolumn[
  \begin{@twocolumnfalse}
\noindent\LARGE{\textbf{Top-seeded solution growth of SrTiO$_3$ crystals and phase diagram studies in the SrO--TiO$_2$ system
}}
\vspace{0.6cm}

\noindent\large{\textbf{Christo Guguschev,$^{\ast}$ Detlef Klimm, Frank Langhans, Zbigniew Galazka, Dirk Kok, Uta Juda, and Reinhard Uecker}}\vspace{0.5cm}

\noindent\textit{\small{\textbf{Received 8th October 2013, Accepted Xth XXXXXXXXX 20XX\newline
First published on the web Xth XXXXXXXXXX 200X}}}

\noindent \textbf{\small{DOI: 10.1039/b000000x}}
\vspace{0.6cm}

\noindent \normalsize{The TiO$_2$ rich part of the $(1-x)$ SrO + $x$ TiO$_2$ phase diagram $0.5\leq x\leq1.0$ was redetermined and the eutectic point between SrTiO$_3$ and TiO$_2$ was found at $x_\mathrm{eut}=0.7700\pm0.0001; T_\mathrm{eut}=(1449\pm3)^{\,\circ}$C. From TiO$_2$ rich melt solutions, $x=0.75$ centimeter-sized single crystals could be grown. The best crystals with etch pit density $<2\times10^4$/cm$^2$ were obtained for growth directions $\left\langle110\right\rangle$ and $\left\langle100\right\rangle$. AFM investigation of the interface reveals layer-by-layer growth.}
\vspace{0.5cm}
 \end{@twocolumnfalse}
  ]

\footnotetext{Leibniz Institute for Crystal Growth, Max-Born-Str. 2, D-12489 Berlin, Germany. E-mail: christo.guguschev@ikz-berlin.de; Fax: +49-30-6392-3003; Tel.:  +49-30-6392-3124}

\section{Introduction}

Tausonite (SrTiO$_3$, STO) is a cubic perovskite and high quality single crystalline bulk crystals are attractive as substrates for oxide heterostructures. Two dimensional electron gases have been found at the interface between SrTiO$_3$ and the following oxides: LaTiO$_3$ \cite{Ohtomo02}, LaAlO$_3$ \cite{Ohtomo04}, LaVO$_3$ \cite{Hotta07}, LaGaO$_3$ \cite{Perna10}, GdTiO$_3$ \cite{Moetakef11}, and KTaO$_3$ \cite{Kalabukhov07}. It also has been found in SrTiO$_3$/SrTi$_{0.8}$Nb$_{0.2}$O$_3$ superlattices \cite{Hwang04}.

The observation of the conducting interfaces has led to an increasing demand of high quality single crystals for fundamental research and devices having extraordinary properties and functionality. Most tausonite substrates which are used nowadays are grown by flame fusion growth with high dislocation densities ($>1\times10^6/\mathrm{cm}^2$). These Verneuil-grown crystals are produced at industrial scale and they are commercially available. When it comes to higher structural perfection only a few crystal growth methods are suitable to reach the high quality demands with a sufficient crystal size. One of the methods is optical floating zone which yields crystals with dislocation densities typically between $(1-5)\times10^5/\mathrm{cm}^2$~~~\cite{Nabokin03}. With a cold crucible Czochralski technique (cold crucible pulling technique) \cite{Aleksandrov81} etch pit densities of $10^2-10^5/\mathrm{cm}^2$ can be reached. SrTiO$_3$ with the lowest dislocation density can be grown from the flux ($0-10^2/\mathrm{cm}^2$) \cite{Rytz90,Scheel76}.

SrTiO$_3$ and TiO$_2$ are setting up a eutectic subsystem where top-seeded solution growth (TSSG) can be used for the growth of SrTiO$_3$ crystals at substantially lower temperature, compared to the growth from stoichiometric melts. However, there is a lack of thermodynamic data at this composition range in the literature. This work highlights the self-flux growth of SrTiO$_3$ single crystals in a cylindrical shape, the refinement of the SrO--TiO$_2$ phase diagram, which now includes the thermodynamic data of the melt, by means of DTA/DSC measurements and a description of typical growth surface morphologies when using different seed orientations by the AFM method.

\section{Experimental}

\begin{figure}[ht]
\centering
\includegraphics[width = 0.48\textwidth]{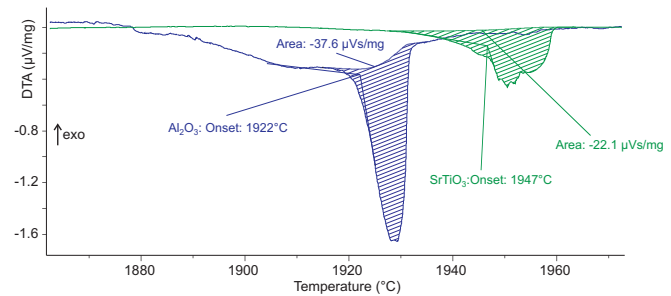}
\caption{DTA heating curves of 80.99\,mg crystalline SrTiO$_3$ and Al$_2$O$_3$ powder standard. The latter melts at $2054^{\,\circ}$C with $\Delta H_\mathrm{f} = 1161$\,J/g \cite{FactSage64}.}
\label{fig:DTA-Xtal}
\end{figure}

Thermal analysis of commercial SrTiO$_3$ crystals (CrysTec GmbH, Berlin) was performed up to the melting point beyond $2000^{\,\circ}$C in a NETZSCH STA 429 calorimeter. Lidded tungsten crucibles in 99.9999\% pure static He atmosphere and W-W/Re thermocouples were used. The significant aging of these thermocouples that is shown under given conditions was accounted for by a subsequent calibration run with Al$_2$O$_3$ powder, where melting point $T_\mathrm{f}$ and heat of fusion $\Delta H_\mathrm{f}$ are known (Fig.~\ref{fig:DTA-Xtal}).

In addition, 23 different $(1-x)$ SrO + $x$ TiO$_2$ powder mixtures with $0.4014\leq x\leq0.9284$ were measured by DTA in a NETZSCH STA 449 calorimeter in flowing oxygen with platinum crucibles. According to the literature (Figs.~297, 298, 2334 in Ref.~\cite{ACer34}) a eutectic point of SrTiO$_3$/TiO$_2$ should be near $x=0.8; 1440^{\,\circ}$C. It was the aim of these measurements to determine this eutectic point, and if possible also the liquidus points close to it, with high accuracy. Such high accuracy can be obtained only with the STA 449 due to lower temperatures. Moreover, the oxygen atmosphere was expected to suppress evaporation especially of strontium.

A solution composition of 75\,mol\% TiO$_2$ and 25\,mol\% SrO was used for the self-flux growth experiments. For these experiments, the dried, mixed and pressed starting powders of SrCO$_3$ and TiO$_2$ with purities of 99.99\%, were heated in platinum crucibles in MoSi$_2$ muffle furnaces at maximum temperatures of $1620^{\,\circ}$C.

\begin{table*}[htb]
\small
  \caption{\ Crystal growth conditions}
  \label{tab:growth}
  \begin{tabular*}{0.60\textwidth}{@{\extracolsep{\fill}}cccccc}
    \hline
    growth  & afterheater & axial     & pulling rate & seed                           & rotation rate \\
  condition & in use      & gradient  & (mm/h)       & orientation                    & (rpm)  \\
    \hline
    T1      & no          & very high & 0.2          & $\left\langle100\right\rangle$ & 20  \\
    T2      & yes         & very low  & 0.2--0.6     & $\left\langle100\right\rangle$ & 20  \\
    T3      & yes         & moderate  & 0.2          & $\left\langle110\right\rangle$ & 5   \\
    T4      & yes         & moderate  & 0.2          & $\left\langle111\right\rangle$ & 5   \\
    \hline
  \end{tabular*}
\end{table*}

Crystal growth was performed in a conventional \emph{rf}-heated Czochralski setup (Cyberstar) with automatic diameter control. Verneuil-grown SrTiO$_3$ single-crystals seeds (supplied by Crystec GmbH, Berlin) with orientations of $\left\langle100\right\rangle$, $\left\langle110\right\rangle$ and $\left\langle111\right\rangle$ were used. The growth runs were carried out in air with use of a platinum crucible embedded in ZrO$_2$ and Al$_2$O$_3$ insulation. Most of the growth runs were performed using a pulling rate of 0.2\,mm/h and with an additional actively heated Pt after heater on top of the crucible. Different growth conditions were tested, which are shown in table~\ref{tab:growth}. The surface temperature of the melt solution at the beginning of the growth was kept around $1540^{\,\circ}$C, which was measured with a two colour pyrometer. This measured value is close to the calculated $T_\mathrm{liq}=1535^{\,\circ}$C (Fig.~\ref{fig:fugacity}).

For EPD measurements samples of the grown crystals were oriented in the $\left\langle100\right\rangle$ direction, chemo-mechanically polished and etched for several minutes in a mixture of 1\,HF-2\,HNO$_3$-2\,H$_2$O \cite{Rhodes66}. A MFD-3D Stand Alone AFM (Atomic Force Microscope) from Asylum Research with an AEK 2002 acoustic isolation enclosure was used to investigate facets which were present at the growth interface. The measurements were done in contact mode in air and at room temperature using standard cantilevers from $\mu$mash (silicon, tip diameter $<10$\,nm, spring constant 2.8\,N/m). The scan size varied between $5\times5\,\mu\mathrm{m}^2$ and $10\times10\,\mu\mathrm{m}^2$.

\section{Results and discussion}

\subsection{Crystal growth and characterization}

\begin{figure}[htb]
\centering
\includegraphics[width = 0.46\textwidth]{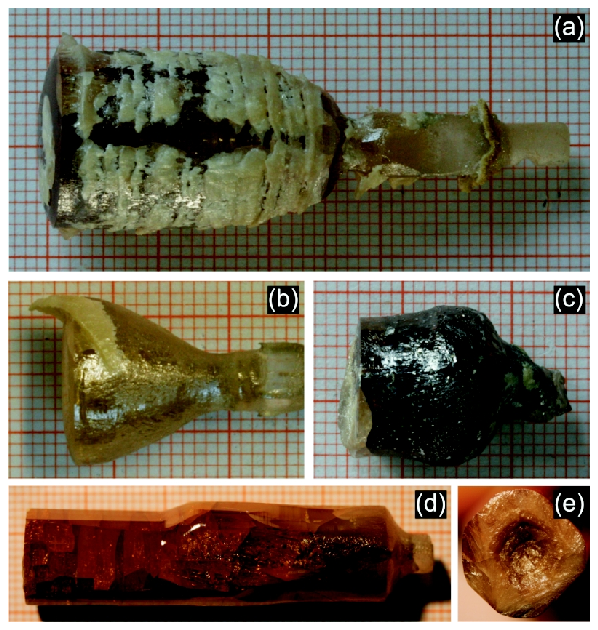}
\caption{SrTiO$_3$ crystals grown by self flux method with convex (a), W-shaped (b), nearly flat (c) and pronounced concave growth interfaces (d,e). Pulling directions of the crystals (a-d) were $\left\langle100\right\rangle$, $\left\langle110\right\rangle$, $\left\langle111\right\rangle$ and $\left\langle100\right\rangle$, respectively and the growth conditions (see table~\ref{tab:growth}) were T1, T3, T4, T2, respectively. Interfaces are partially covered by the frozen melt solution (a--e) and by crystals grown from the vapor phase (a). A pulling rate of 0.2\,mm/h was used for the crystals shown in a--c and mainly 0.35\,mm/h was used for the crystal shown in d and e.}
\label{fig:Xtals}
\end{figure}

By using an inductively heated Czochralski setup cylindrical SrTiO$_3$ single crystals could be obtained. In Fig.~\ref{fig:Xtals}a a single crystal is shown, which was pulled in the $\left\langle100\right\rangle$ direction. Even by using Verneuil-grown seeds, high quality tausonite crystals with a representative EPD value of $1.7\times10^4/\mathrm{cm}^2$ have been grown. The convex growth interface left in Fig.~\ref{fig:Xtals}a shows a (100) facet at its center. At the cylindrical part of the crystal, $\lbrace110\rbrace$ facets, which grow slowest, are present. Mainly the non-faceted faces are covered by rutile (TiO$_2$) and tausonite crystals grown from the vapour phase. To minimize the condensation of rutile on the surface of single crystals, the application of an after heater is necessary which keeps the oversaturation of titanium oxide in the vapour as low as possible.

Growth interfaces between nearly flat (Fig.~\ref{fig:Xtals}c), w-shaped (Fig.~\ref{fig:Xtals}b) and concave (Fig.~\ref{fig:Xtals}d--e) were observed, depending on the growth conditions (table~\ref{tab:growth}) and orientation of the crystals. Non-convex interfaces, especially strong pronounced concave growth interfaces (Fig.~\ref{fig:Xtals}e) lead to a formation of cavities in the central region of the crystal, where the melt can penetrate into the crystals core, causing the crystal to crack upon cooling, which can be seen in Fig.~\ref{fig:Xtals}d. The outer parts of the crystal retain the high crystal quality (EPD $1.5\times10^4/\mathrm{cm}^2$), which is similar to the crystals pulled in $\left\langle100\right\rangle$ direction without use of an additional after heater. The cracking can be explained by the deviations of thermal expansion properties of the two solid phases, tausonite and rutile, which crystallize simultaneously from the melt upon reaching the eutectic temperature. The difference of the thermal expansion coefficient is more than one order of magnitude \cite{Ligny96,Kirby67}. Moreover, a concave interface leads to a higher density of grown-in dislocations which also may cause cracking. Appropriate growth and post-growth conditions have to be applied for different pulling directions to yield convex growth interfaces, prevent condensation of rutile at crystal surfaces and reduce the amount of frozen melt solution at the bottom of the crystal at the same time.

\begin{figure}[htb]
\centering
\includegraphics[width = 0.48\textwidth]{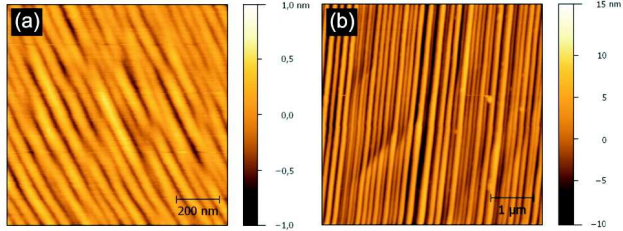}
\caption{AFM measurements of the as-grown facets at the central part of the melt-solid interfaces for different pulling directions. (a) $\left\langle100\right\rangle$ and (b) $\left\langle110\right\rangle$ direction.}
\label{fig:afm1}
\end{figure}

\begin{figure}[htb]
\centering
\includegraphics[width = 0.48\textwidth]{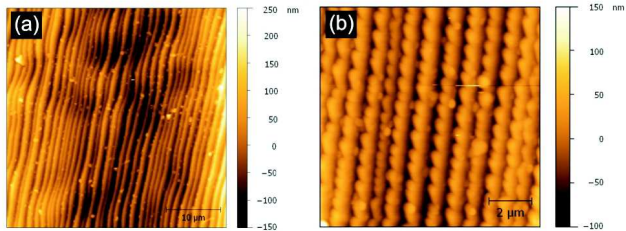}
\caption{AFM measurements of the as-grown $\lbrace111\rbrace$ facet at outer (a) and central (b) part of the melt-solid interface.}
\label{fig:afm2}
\end{figure}

Figs.~\ref{fig:afm1} and \ref{fig:afm2} show AFM pictures taken on the different growth facets. The measurements were performed at melt droplet free regions. Step flow can be observed for the three surfaces with different step heights. On the one hand, step bunching is indicated in Fig.~\ref{fig:afm1}b for the $\left\langle110\right\rangle$ pulling direction with step heights between 5 and 31 unit cell distances $a$ and average step distances of $136\pm40$\,nm. On the other hand, step heights between $1-2a$ and step distances of $65\pm12$\,nm were measured at the $\lbrace100\rbrace$ facet in case of the pure step flow (Fig.~\ref{fig:afm1}a). Furthermore, we observed step meandering instabilities at $\lbrace111\rbrace$ surfaces, which can be explained by the Bales-Zangwill mechanism \cite{Bales90}. Figs.~\ref{fig:afm2} shows steps at the $\lbrace111\rbrace$ surface with step heights between 44 and $418a$ and step distances of $1115\pm322$\,nm. The observed differences of the surface morphologies are attributed to the difference of the surface energies which increase in the order $\lbrace100\rbrace$, $\lbrace110\rbrace$, $\lbrace111\rbrace$ \cite{Sano03}. Investigations of the dislocation densities of the grown crystals have shown, that the dislocation density is similar for crystals which were pulled in $\left\langle100\right\rangle$ and $\left\langle110\right\rangle$ direction (growth condition T1--T3, EPD ca. $(1-2)\times10^4$/cm$^2$). In contrast to that, a higher dislocation density ($8\times10^4/\mathrm{cm}^2$) was measured for the crystal which was pulled in $\left\langle111\right\rangle$ direction (T4). These results indicate that the $\left\langle110\right\rangle$ and $\left\langle100\right\rangle$ pulling directions are more favourable due to stable step flow growth and lower dislocation densities.

\subsection{Thermal analysis}

SrTiO$_3$ shows one melting peak (Fig.~\ref{fig:DTA-Xtal}), also during repeated DTA runs, which indicates congruent melting. If the measured onset temperature for SrTiO$_3$ is compared to the Al$_2$O$_3$ standard, a fusion point 
$T_\mathrm{f}=(2054-1922)^{\,\circ}\mathrm{C} + 1947^{\,\circ}\mathrm{C} = (2079\pm20)^{\circ}\mathrm{C}$
is obtained. The error is estimated from the fluctuation of results for these and other substances in this temperature range. Even if a DSC is not available for such high $T$, the heat of fusion can be estimated by comparing the peak area with that of the standard. (One can assume that the sensitivity of the sample carrier is almost identical for SrTiO$_3$ and Al$_2$O$_3$ because their fusion points differ by only 25\,K.) One finds $\Delta H_\mathrm{f} \approx 683$\,J/g = 125.2\,kJ/mol. Here a larger error of ca. $\pm20$\% must be assumed, derived from the insufficient reproducibility of this determination. At least this value seems not unrealistic: for the structurally similar CaTiO$_3$ the FactSage \cite{FactSage64} database gives at its melting point $1960^{\,\circ}$C a $\Delta H_\mathrm{f} = 106.638$\,kJ/mol.

\begin{figure}[ht]
\centering
\includegraphics[width = 0.46\textwidth]{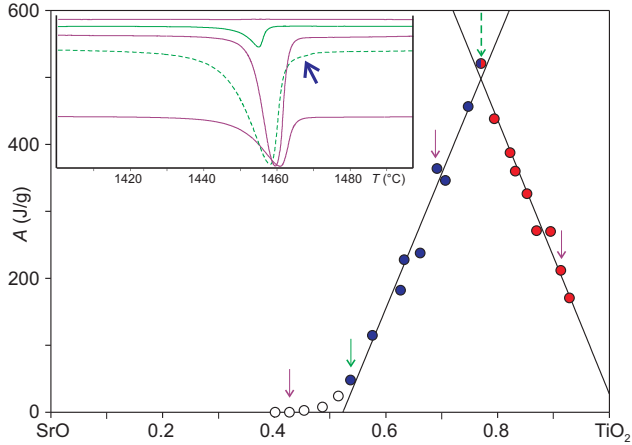}
\caption{Tammann plot of the eutectic peak area $A$ vs. TiO$_2$ mole fraction $x$. Full points left or right from $x_\mathrm{eut}$ were fitted by linear functions separately. A few DTA curves of the eutectic peaks are shown in the insert; the corresponding points in the plot are marked by an arrow. The dashed DTA curve is for $x=0.7700$, which is almost the intersection of the linear fits at $x_\mathrm{eut}=0.7701$.}
\label{fig:DTA-mix}
\end{figure}

For the $(1-x)$ SrO + $x$ TiO$_2$ powder mixtures coming from the SrO rich side, $x=0.4270$ was the last composition not showing any sign of a eutectic peak (uppermost DTA curve in the insert of Fig.~\ref{fig:DTA-mix}), and the next composition, $x=0.4533$ ($3^\mathrm{rd}$ experimental point from the left in Fig.~\ref{fig:DTA-mix}), already showed a minor eutectic peak ($A=2.6$\,J/g). The largest eutectic peak was found for $x=0.7700$ ($A=520.7$\,J/g). The extrapolated onsets of the eutectic peaks vary for all compositions in a small range of $(1449\pm3)^{\,\circ}$C, which is the eutectic temperature $T_\mathrm{eut}$. It is obvious from the lever rule that the peak area $A$ varies linearly with $x$, with a maximum at $x_\mathrm{eut}$. Fig.~\ref{fig:DTA-mix} shows such Tammann construction and from the intersection of linear fits $A(x)$ from both sides one obtains $x_\mathrm{eut}=0.7701$, which almost coincides with the experimental point and the DTA curve that is marked by dashed lines.

\subsection{Thermodynamic calculations}

With the Schr\"oder-van Laar equation one can calculate a pseudo-binary phase diagram SrTiO$_3$--TiO$_2$ from the melting point and heat of fusion of both end members. Data for SrTiO$_3$ were estimated by the DTA measurements shown in Fig.~\ref{fig:DTA-Xtal} and reliable values for TiO$_2$ ($T_\mathrm{f}=1857^{\,\circ}$C, $\Delta H_\mathrm{f}=46024$\,J/mol) can be found in FactSage databases \cite{FactSage64}. With these data a hypothetical eutectic point at $x=0.8053$ (in scales of SrO/TiO$_2$) and $T=1652^{\,\circ}$C can be calculated if the melt was ideal, which is far way from the measured values $x=0.7701, T_\mathrm{eut}=1449^{\,\circ}$C. To reach such low $T_\mathrm{eut}$ for an ideal melt, $\Delta H_\mathrm{f}$ of SrTiO$_3$ would have to be only $\frac{1}{3}$ of its measured value, which can be ruled out. Instead one can conclude that the melt is stabilized by a very large negative excess enthalpy, probably resulting from the strong chemical interaction between the Lewis base SrO and the Lewis acid TiO$_2$.

\begin{table}[htb]
\small
  \caption{\ Assessed parameters for the description of melts from $0.5\leq x\leq1.0$ by (\ref{eq:G})}
  \label{tab:asse}
  \begin{tabular*}{0.40\textwidth}{@{\extracolsep{\fill}}rrrr}
    \hline
   $a$ (TiO$_2$) & $b$ (SrO) & $A$       & $B$      \\
    \hline
   1             & 1         & -256452   & 44.6683  \\
   1             & 2         & -7038.11  & -19.2816  \\
   1             & 4         & -1042.50  &  7.01178  \\
   2             & 1         & 301.955   & -16.3740  \\
   4             & 1         & 1385.40   &  32.9610  \\
    \hline
  \end{tabular*}
\end{table}

The excess Gibbs energy of the melt was expressed by
\begin{equation}
\sum_{a,b} G_\mathrm{ex} = y^bx^a(A+BT)  \label{eq:G}
\end{equation}
where $y=1-x$ is the molar fraction of SrO. Reasonable fits of the experimental data ($T_\mathrm{f}$ for SrTiO$_3$, $T_\mathrm{eut}$ and $x_\mathrm{eut}$, a few liquidus temperatures for compositions near $x_\mathrm{eut}$) were obtained only if exponents $a,b$ up to 4 were taken into consideration. The assessment was performed with FactSage \cite{FactSage64} for compositions between SrTiO$_3$ and TiO$_2$ only, because only there experimental data were available. Results are given in table~\ref{tab:asse}. Even if not used in the assessment itself, it turned out the heat of fusion for SrTiO$_3$ that can be calculated from the difference of the enthalpy of the 1:1 composition directly below and above $T_\mathrm{f}$ (126.24\,kJ/mol) corresponds almost perfectly with the experimental value mentioned above ($\Delta H_\mathrm{f}=125.2$\,kJ/mol). Only Gibbs energy data for the melt were assessed here, and data for solid SrTiO$_3$ were taken from recent accurate galvanic measurements: $\Delta H^0_\mathrm{f}=-1653.290$\,kJ/mol, $S=109.128$\,J/(mol$\cdot$K), $c_p=137.09676 + 0.00323T - 456.7919tT^{-0.5} - 1195220T^{-2}$ (in J/(mol$\cdot$K) \cite{Jacob11}.

\begin{figure}[htb]
\centering
\includegraphics[width = 0.42\textwidth]{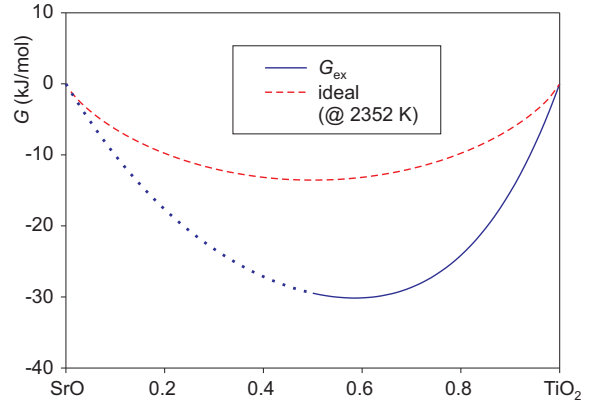}
\caption{Gibbs free energy of liquid SrO/TiO$_2$ mixtures at the melting temperature of SrTiO$_3$. Solid line: real data (equation (\ref{eq:G}) and table~\ref{tab:asse}), dashed line: ideal mixture for comparison. $G_\mathrm{ex}$ data for $x<0.5$ are dotted because no experimental results were available.}
\label{fig:G_ex}
\end{figure}

Fig.~\ref{fig:G_ex} demonstrates that indeed the liquid phase is considerably stabilized, compared with an assumed ideal mixture (dashed curve) at $T_\mathrm{f}$ of SrTiO$_3$. The asymmetry of the $G_\mathrm{ex}(x)$ curve is responsible for the position of the eutectic point near $x=0.77$ that would otherwise, for a symmetric thermodynamic potential, be shifted to larger $x>0.8$.

\section{Discussion and conclusions}

\begin{figure}[htb]
\centering
\includegraphics[width = 0.46\textwidth]{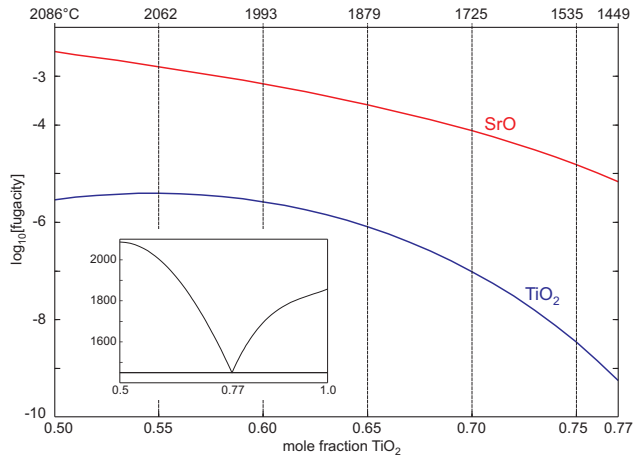}
\caption{Fugacity of the main species SrO(g) and TiO$_2$(g) over a $(1-x)$ SrO + $x$ TiO$_2$ melt for compositions ranging from SrTiO$_3$ ($x=0.5$) to the SrTiO$_3$/TiO$_2$ eutectic ($x=0.77$) at the corresponding liquidus temperatures $T_\mathrm{liq}(x)$ which are given as labels on the upper abscissa. The insert shows the phase diagram between SrTiO$_3$ and TiO$_2$.}
\label{fig:fugacity}
\end{figure}

With the assessed thermodynamic data, equilibria between solids (SrTiO$_3$, TiO$_2$), liquids (SrO--TiO$_2$ melt with $G_\mathrm{ex}$ described by (\ref{eq:G})), and gas phase (SrO(g), TiO$_2$(g)) can be calculated, and the results are summarized in Fig.~\ref{fig:fugacity}. It turns out that SrO has the highest vapour pressure, expressed by its fugacity $f$. $p_\mathrm{SrO}$ is $\gtrsim3$\,mbar for pure SrTiO$_3$ at its melting point. Such high fugacity can lead to strontium loss of the melt, which could be accounted for by a slight SrO excess in the feedstock. Then, however, excess SrO might be dissolved in the solid e.g. under the formation of Ruddlesden-Popper (RP) phases impeding crystal quality \cite{Tilley77,Ruddlesden58}. Measurements with polycrystalline SrO(SrTiO$_3$)$_n$ up to 1000\,K showed that the thermal conductivity $\kappa$ is maximum for the cubic perovskite SrTiO$_3$ ($n=\infty$) and significant lower for RP phases $n=1$ (Sr$_2$TiO$_4$) and $n=2$ (Sr$_3$Ti$_2$O$_7$) \cite{Wang08}. Oh et al. \cite{Oh11} pointed out that phonons with wavelength in the order of 1\,nm dominate heat transport. Sr$_2$TiO$_4$ has perpendicular to the SrO interlayers a lattice parameter $c=1.259$\,nm (cf. Fig.~3 top in Ref. \cite{Lee06}) and one can assume that phonon scattering at the SrO interlayers is responsible for the drop of $\kappa$. For crystal growth from the melt, too low $\kappa$ leads to insufficient heat flux through the crystal, and thus to unstable growth.

\balance

By using the TSSG method and TiO$_2$ rich melts, cylindrical SrTiO$_3$ single crystals were grown. In this study a starting composition $x_0=0.75$ was used, which is close to $x_\mathrm{eut}=0.77$. The proximity of $x_0$ to $x_\mathrm{eut}=0.77$ restricts crystal yield, but keeps strontium loss by SrO evaporation negligible ($p_\mathrm{SrO}=1.5\times10^{-5}$\,bar at $T_\mathrm{liq}=1535^{\,\circ}$C, Fig.~\ref{fig:fugacity}). Under the given experimental conditions (Pt crucible in air) $x_0$ could be shifted only slightly towards SrTiO$_3$ to increase yield, because then the liquidus temperature quickly exceeds the stability limit of platinum (e.g. $T_\mathrm{liq}=1725^{\,\circ}$C with $p_\mathrm{SrO}=7.5\times10^{-5}$\,bar at $x=0.70$). 

In this study, even by using Verneuil-grown seeds, tausonite crystals with higher structural perfection were obtained. Special attention is necessary to avoid the growth of rutile (TiO$_2$) on the grown crystal surfaces, and to keep a convex growth interface at the same time. The avoidance of rutile growth and the minimization of frozen melt solution droplets at the growth interfaces reduce the risk of crystal cracking during cooling to room temperature.

AFM investigations have shown that the growth is dominated by the step flow growth mode. Pure step flow, step bunching and step meandering instabilities were observed at growth interface facets $\lbrace100\rbrace$, $\lbrace110\rbrace$ and $\lbrace111\rbrace$, respectively. It can be concluded that $\left\langle110\right\rangle$ and $\left\langle100\right\rangle$ pulling directions from TiO$_2$ rich melts are favourable at low pulling speeds (0.2\,mm/h), to establish stable step flow growth modes at the growth facets and low dislocation densities in the crystals.




\footnotesize{

\end{document}